\documentclass[fleqn,usenatbib]{mnras}
\usepackage{newtxtext,newtxmath}
\usepackage[T1]{fontenc}
\DeclareRobustCommand{\VAN}[3]{#2}
\let\VANthebibliography\thebibliography
\def\thebibliography{\DeclareRobustCommand{\VAN}[3]{##3}\VANthebibliography}
\usepackage{graphicx}	
\usepackage{amsmath}	
\usepackage{commath}
\usepackage{orcidlink}

\title[Observational constraints on complex quintessence]{Observational constraints on complex quintessence with attractive self-interaction}

\author[B. Carvente et al.]{
Belen Carvente\orcidlink{0000-0002-7050-7769},$^{1}$\thanks{E-mail: belen.carvente@correo.nucleares.unam.mx}
V\'ictor Jaramillo\orcidlink{0000-0002-3235-4562},$^{1}$
Celia Escamilla-Rivera\orcidlink{0000-0002-8929-250X}$^{1}$
and Dar\'{\i}o N\'u\~nez\orcidlink{0000-0003-0295-0053}$^{1}$
\\
$^{1}$Instituto de Ciencias Nucleares, Universidad Nacional
  Aut\'onoma de M\'exico, Circuito Exterior C.U., A.P. 70-543,
  M\'exico D.F. 04510, M\'exico
}

\date{Accepted XXX. Received YYY; in original form ZZZ}
\pubyear{2020}

\begin{document}
\label{firstpage}
\pagerange{\pageref{firstpage}--\pageref{lastpage}}
\maketitle


\begin{abstract}

In this paper we consider that dark energy could be described solely by a complex scalar field with a Bose-Einstein condensate-like potential (denoted as \textit{CSFDE}), that is, with a self-interaction and a mass term. In particular, we analyse a solution which in a fast oscillation regime at late-times behaves as a Cosmological Constant. 
Our proposal adequately describes the standard homogeneous and flat Fridman dynamics,  furthermore, in this quintessence--complex scalar field scenario it is possible to mimic the dynamics related to dark energy.  
However, when the precision cosmological tests are implemented in this landscape, the generic Equation-of-State derived for this model in a restricted regime of $a_i$ (which corresponds to the scale factor at which the scalar field turns on), cannot be constrained by late-time current observations, since the analysis constraints solely the scalar field parameters within values ruled out by the theoretical model. This result is a \textit{clear} hint to consider future CSFDE models with, for instance, two scalar fields in order to study the early-time dynamics of the Universe.

\end{abstract}



\section{Introduction}
The inclusion of the dark components of the Universe in Einstein equations gives a consistent description of the current observed dynamics at a large scale \citep{Peebles:1994xt, Ostriker:1995rn, Dodelson:2003ft}. This components are known as dark energy and, at a galactic level \citep{1983ApJ...267..465D, 1974ApJ...193L...1O, 2008ASPC..395..283R, 2000PASP..112..747R, 2018RvMP...90d5002B}, dark matter.
The nature of such
dark components remains unknown. 
Dark energy, 
although at first order is modelled as a repulsive gravitational term such as the Cosmological Constant $\Lambda$, certain observations have shown some tensions in the Hubble flow in the standard $\Lambda$CDM model \citep{2017NatAs...1..627Z}, so it seems like it is not sufficient to describe the dark energy with a constant term; it is thus proposed to be modelled by 
different types of matter,
such that the relation between the spatial components of the corresponding stress energy tensor to the temporal one, is consistent with the observed dynamics; that is, using an analogy with fluid dynamics, it can be defined an Equation-of- State (EoS) $T^i_i=-w\,c^2\,T^0_0$ ({\it i. e.} $p=w\,\rho$ for the pressure and density of a fluid like description). The behaviour of the function $w$ can be related to the observations, as described bellow, and its value at the present is close to minus one. 

Regarding the modelling of the dark matter, several models have proposed that it should be considered as a weakly interactive particle. However, not strong evidence of such a particle has been detected in the current projects that have been created {\it ex professo} to obtain a detection either directly \citep{2013EPJC...73.2648B, 2015arXiv150400820D, 2016EPJC...76...25A, 2017PhRvL.119r1302C} or indirectly \citep{2011ICRC....5..141D, 2015JCAP...09..008F, 2020NuPhA.99621712Y}. It must be faced the possibility that dark matter had zero interaction with the baryonic matter.

Indeed, as mentioned above, the Theory of General Relativity allows to describe several kinds of matter/energy, in comparison to the Newtonian case. In this way, once there are models of one type of matter or another, consistent with the observations, the next step is the determination of characteristic features generated on the baryonic matter by each kind of matter which could, in principle,
be detected. Even supposing that there is no interaction of the baryonic matter with the dark components, other than gravitational, 
it can still be seen that the density distribution of the different types of matter/energy has a very distinctive 
feature which affects the distribution of baryonic matter, and that could tell at least what kind of matter better describes the observed density distribution, see e.g \citet{Nunez:2014fxa} for a discussion on the subject. As an example of the latter, in
 \citet{Dominguez-Fernandez:2017nxx} was studied how the density perturbations evolve inside a dark matter halo considering that the matter was a collection of non-interactive particles, whose dynamics are described by the Vlasov equation. The main result in that work was that the final state has a very distinct distribution in the coordinate space, a double peaked Gaussian in the density and, in the phase space, a volcano-like form in the distribution function, features that could affect the baryonic density distribution, which is an observable quantity.
 
To define analogous strategies regarding the dark energy, there are 
several considerations that must be taken into account. 
It is a component associated mainly with the cosmic acceleration 
\citep{1998AJ....116.1009R, 2000PhST...85...47G, 2016A&A...594A..14P} which, as mentioned above, 
can be modelled with the simple inclusion of a properly tuned $\Lambda$
in the Einstein's equations, although this constant rules out
the usual Minkowski's solution, and the asymptotic limits of all the well established solutions to the Einstein's equations need to be modified.
It is an exciting fact that there is a new constant of Nature, see \citet{Bianchi:2010uw} for an interesting discussion on the subject, but the implications in the equations themselves enhance the need to prove the veracity of such 
model, a fact which is done proposing more general models to describe the cosmic acceleration. 
In addition to its modeling with different kinds of matter, another way to proceed, is to propose alternative gravity
models of matter that can describe the current observed dynamics \citep{Clifton:2011jh,Jaime:2018ftn,Escamilla-Rivera:2019hqt}.

Within the models proposed to describe the dark energy other than a constant term, in Einstein gravity, those considering a scalar field can be the simplest, well motivated choice from a particle physics point of view. Nevertheless, the great challenge is to determine the appropriate scalar potential $V(\Phi)$ that could explain current cosmological observations. An example of a description by a scalar field minimally coupled to gravity is the \textit{quintessence} model. The main motivation for considering it is to reduce the so-called \textit{fine--tuning} problem, issue that has been explored by the tracker quintessence solutions. However, the predicted values on these models for the EoS at the present epoch is not in good agreement with supernovae results \citep{Zlatev:1998tr, Steinhardt:1999nw}. Another example is the exponential quintessence potential that focus on models and parameters which lead to inflation, nevertheless nucleosynthesis constraints require that the energy density of the scalar field be $\Omega_\phi \leq 0.2$, i.e., it would never dominate the Universe \citep{Ferreira:1997hj}. Another dynamical potentials proposed in \citet{Sahni:1999qe} and \citet{UrenaLopez:2000aj} avoid successfully the fine--tuning and cosmic coincidence problem, but the values of the potential parameters can not be unambiguously determined in order to  match the observations constraints.

The above models are made up of real scalar fields. However, also complex scalar fields should be considered since such fields (unlike the real case) have been invoked in many different sectors of particle physics \citep{Gu:2001tr} (such as the Higgs mechanism) and interestingly in the scene of ultra cold gases \citep{RevModPhys.74.875}; they can be used to construct static distributions as Boson stars \citep{Ruffini69}, and also configurations surrounding a black hole, the so--called wigs \citep{Burt:2011pv}, and they
can even define static configurations with an associated angular momentum number \citep{Alcubierre:2018ahf, Carvente:2019gkd}. Furthermore, a real quantized scalar 
field yields the same field equations as those obtained by using a classical complex scalar field \citep{Barranco:2010ib}. These reasons motivated us to consider a dark energy model described by a massive quintessence--complex scalar field with attractive self interaction. Such field was formerly studied in \citet{Suarez:2016eez} and, in the present work, we revisited the idea focusing in the so-called peculiar branch solution of the Einstein-Klein-Gordon equations in order to obtain parameter restrictions of the potential consistent with the current precision observations. Although we are aware of the latest results regarding the possible dynamical behavior of the EoS \citep{2017NatAs...1..627Z} and the impossibility for a single canonical field to evolve crossing over $w=-1$ because of the no-go theorem \citep{quintomcai}, it is interesting to explore in detail the properties of the previously mentioned branch and in computing best fit values of their parameters, in order to have a quantitative description of the model and a clearer picture of what the model needs in order to be consistent with such a dynamical behavior of the dark energy.

This paper is organized as follows: 
in Sec.~\ref{sec:complexfield} we briefly describe the Fridman (we use the direct transliteration from the Russian) background considering a complex scalar field instead of the 
standard cosmological constant.
In Sec.~\ref{sec:quintessence} we study the Einstein-Klein-Gordon equations to describe dark energy based in the fact that the scalar potential can be proposed as an effective \textit{fluid}, with the caution of not solving the EoS, but solving Klein--Gordon first, and with the field and its derivative, compute the density and the scalar pressure, and subsequently compute the corresponding $w$. We
consider the fast oscillation regime, where the pulsation $\omega$ of the scalar field is assumed to be faster than the Hubble expansion.
An EoS is obtained for a peculiar branch in such fast oscillation regime. We denote the model presented in this manuscript as Complex Scalar Field Dark Energy (CSFDE). In Sec.~\ref{sec:eos_scalar} a generic EoS with a complex scalar field mimicking the dark energy term is presented. A description of the current late-time observations are given in  Sec.~\ref{sec:surveys}. These samplers will be employed to constrain the \textit{only} free cosmological parameter that goes into the expression for the EoS of our CSFDE model. In Sec.~\ref{sec:methodology} we describe the methodology to proceed with the precision analysis and discuss the cosmological constraints obtained. 
Finally, our conclusions are given in Sec.~\ref{sec:conclusions}. 

\section{Complex Scalar Field in an homogeneous background}
\label{sec:complexfield}

In this section we derive first the evolution equations for a homogeneous and flat universe filled with radiation, baryonic and dark matter components and an effective density which will mimic the dark energy component. In the second part, we introduce the complex scalar field to describe such effective density and obtain the corresponding Klein-Gordon equation.

\subsection{Fridman equations}
\label{sec:fridman}

First, let us consider a homogeneous isotropic Universe, described by the
Fridman-Lem\^aitre metric
\begin{equation}
ds^2=-c^2dt^2+a^2(t)\left[\frac{dr^2}{1-Kr^2}+r^2(d\theta^2+\sin^2\theta d\varphi^2)\right], \label{eq:ele_F}
\end{equation}
where $a(t)$ is the scale factor and $K$ the curvature scalar. From this point forward we consider spatial flatness.
As it is standard, we can derive the Fridman equation and the energy conservation equation by introducing the above metric in the Einstein's equations. Before continue with this straightforward calculation, let us establish our pivot model:
the paradigmatic cosmological model, $\Lambda$CDM, which considers a total density of the Universe $\rho_T = \rho_{r} + \rho_b + \rho_{\text{cdm}}+\rho_{\text{DE}}$, normalised by the critical density given by $\rho_{\rm crit}=3 H_0^2/ 8\pi G$, where $H_0$ is the Hubble parameter at present time and $G$ is the gravitational constant. According to this, we can derive the constraint equation from the Fridman evolution as
\begin{equation} \label{eq:h_Fr}
\left(\frac{H}{H_0}\right)^2=\Omega_{m} + \Omega_\Lambda,
\end{equation}
with
\begin{equation}\label{eq:matter_density}
\Omega_{m}=\frac{\Omega_{r,0}}{a^4} + \frac{\Omega_{b,0}}{a^3} + \frac{\Omega_{\text{cdm},0}}{a^3},
\end{equation}
where $\Omega_i=\rho_i/\rho_{\rm crit}$ ($i=\text{cdm},b,r$), represents the density parameter and the symbols $\text{cdm},b,r$ correspond to cold dark matter (CDM), baryonic matter and radiation, respectively. 

According to \citet{Aghanim:2018eyx}, the cosmological values for the densities described above are: $\Omega_{\text{cdm}} h^2=0.120 \pm 0.001$, $\Omega_b h^2 = 0.0224\pm 0.0001$, $\Omega_{\Lambda}= 0.674\pm 0.013$ and $\Omega_{m}=0.315\pm 0.007$.
Currently, this model has proved to be consistent with several observations, however, it has problems in regards to the tension on the value of some parameters like those of $\sigma_8$ and $H_0$ \citep{Verde:2019ivm}.

As indicated in the Introduction, it is interesting to explore dynamic EoS since they alleviate tensions between certain cosmological parameters. Classical scalar fields are simple models for introducing time-dependent equations of state. The case of a 
scalar field minimally coupled to gravity, with a positive canonical kinetic term, called quintessence \citep{Copeland:2006wr,Zlatev:1998tr}. Extensions to this model have been widely considered, for example some by including non-canonical scalar fields or negative signed kinetic terms. However in this work, we consider a simpler case, which considers a rapidly oscillating minimally coupled complex scalar field \citep{Gu:2001tr,Boyle:2001du,Suarez:2016eez}.


\subsection{The Klein-Gordon equation}
\label{sec:kg}

We use the evolution described as a starting point, and introduce a complex scalar field 
in order to model dark energy. Our proposal is based in the fact that the scalar potential $V(\abs{\Phi}^2)$, has a quartic-form with a negative scattering length as

\begin{equation}\label{eq:V}
V(\abs{\Phi}^2)=\frac{m^2c^2}{2\hbar^2}\abs{\Phi}^2-\frac{2\pi A_s m}{\hbar^2}\abs{\Phi}^4,
\end{equation}
where $m$ is the complex scalar field mass, $A_s$ the absolute value of the scattering length and $\hbar$ the reduced Planck constant.
This scalar potential describes, for instance, a relativistic Bose-Einstein condensate at zero temperature with attractive self-interaction \citep{Castellanos:2013ena,Castellanos:2015nbe}, and it is also similar to the Higgs potential of particle physics but with an overall opposite sign.

The evolution of this complex scalar field in the cosmological scenario described above is given by the Klein-Gordon equation
\begin{equation}\label{eq:KG}
\frac{1}{c^2}\frac{d^2\Phi}{dt^2}+\frac{3 H}{c^2}\frac{d\Phi}{dt}+2\frac{d V}{d\abs{\Phi}^2}\Phi=0,
\end{equation}
from where we can express the complex scalar field as
\begin{equation}
\Phi=|\Phi|e^{i\theta}.   \label{def:Phi}
\end{equation}
Solutions to the Einstein-Klein-Gordon equations would require in total six parameters related to initial conditions for the real and imaginary parts of $\Phi$ and their first time derivative together with the scalar field values for $m$ and $A_s$. A further simplification can be made within this model when, consistently with dark energy-like behavior, we assume that the field is oscillating rapidly, this leads to a three-parameter model.

In our proposal, we are going to follow
the procedure given in \citet{Suarez:2016eez}, of which we summarize some key points. Using (\ref{def:Phi}) in (\ref{eq:KG}), 
the Klein-Gordon equation can be divided into a real and an imaginary part, from which the second leads to the equation:

\begin{equation}\label{def:Q}
Q=-\frac{1}{\hbar c^2}a^3\abs{\Phi}^2\frac{d\theta}{dt},
\end{equation}
where $Q$ a is constant\footnote{After integration, the imaginary part of the Klein-Gordon equation leads to a conserved quantity, which corresponds to the conserved charge of a complex scalar field, given by $Q=\frac{1}{c^2\hbar}\int dx^3\sqrt{-g}\ \text{Im}(\Phi\partial_t\Phi^*)$.} and $a$ the scale factor. 

From the real part, and using the conserved charge $Q$ explicitly in this equation, we obtain
\begin{equation}\label{eq:real}
\frac{1}{c^2}\left[\frac{d^2\abs{\Phi}}{dt^2}-\frac{Q^2\hbar^2 c^4}{a^6\abs{\Phi}^3}\right]+\frac{3 H}{c^2}\frac{d\abs{\Phi}}{dt}+2\frac{d V}{d\abs{\Phi}^2}\abs{\Phi}=0.
\end{equation}

The term containing $Q^2$ is usually related to a centrifugal force when making the analogy of this equation with that of fictitious particle with radial coordinate $\abs{\Phi}$, hence the name spintessence for that model \citep{Suarez:2016eez,Boyle:2001du}.

In the real case, with a quartic potential analogous to (\ref{eq:V}), we have $\theta=0$, therefore the conserved quantity $Q$ in (\ref{def:Q}) is equal to zero, implying among other things, that the solution must have a rapidly oscillating behavior with an equation of state also oscillating around $w=0$ \citep{Magana:2012ph} and the solutions to the equation of motion must be obtained by numerical integration in an appropriate set of variables. The quartic potential is not the only possibility, for instance taking a massless scalar field ($\mu=0$, $\lambda=0$) the equation of state stays trivially at the value $w=1$. Other (real) scalar fields, describing quintessence potentials, as the ones listed in the introduction, may have dynamical EoS some of which also oscillate in time. 
In this work we take the opposite approach, namely $Q\gg0$, leading to an exact solution of the problem which is useful in the implementation of tests for the model with cosmological analyzes.
To compute the energy density and pressure of the complex scalar field, we consider the following expressions:

\begin{equation}\label{eq:energia}
\epsilon=\frac{1}{2c^2}\left|\frac{d\Phi}{dt}\right|^2+V(\abs{\Phi}^2),
\end{equation}

\begin{equation}\label{eq:presion}
P=\frac{1}{2c^2}\left|\frac{d\Phi}{dt}\right|^2-V(\abs{\Phi}^2).
\end{equation}
Notice how we can connect these equations to the ones presented in Sec.\ref{sec:fridman}
where
the quantity $\epsilon$ will replace the $\Lambda$CDM quantity $\rho_{\text{crit}}\Omega_\Lambda$ in the Fridman equation.

From the equations (\ref{eq:KG}), (\ref{eq:energia}) and (\ref{eq:presion}) we can obtain a useful equation for the energy density that resembles the continuity equation for a perfect fluid

\begin{equation}\label{eq:paraintegrar}
\frac{d\epsilon}{da}+\frac{3}{a}(\epsilon+P)=0.
\end{equation}

With these equations, now we are ready to study particular solutions of the Einstein-Klein-Gordon system evolving with a complex scalar field mimicking the dark energy component.
As metioned, this particular model in the fast oscillation regime and its homogeneous solution have already been presented previously by \cite{Suarez:2016eez}, and we extend the study in order to obtain analytical expressions for most of the quantities of the solution, including $w(z)$.


\section{Dark Energy Solution in the fast oscillation regime}
\label{sec:quintessence}

In \citet{Suarez:2016eez} was found that in the fast-oscillation regime, 
i.e., when the oscillation frequency of the scalar field is much larger than the value of the Hubble function, the solution of the Einstein-Klein-Gordon equations for the case of a complex scalar field with an attractive self interaction potential (\ref{eq:V}) has two different solutions.
One solution (called normal branch) resembles to a dark matter scalar field, while the other solution (called peculiar branch) corresponds to a quintessence model. This solution only exists in the fast oscillation regime, in which the scalar field suddenly emerges and behaves as dark energy at late times.

Following the same logic, in this paper we propose a deduction of an exact solution for the equation of state of the quintessence field. 
Once with this equation, we explore their possible constraints by using current observational data.


\subsection{Peculiar branch solution in the fast oscillation approximation}

To establish the fast oscillation regime mentioned above, we consider the following condition which needs to be satisfied during the evolution of the scalar field

\begin{equation}\label{eq:cond1}
\omega=\frac{d\theta}{dt}\gg H.
\end{equation}
In addition to the latter condition, we will impose that the magnitude of the scalar field change slowly on time respect to the angular frequency of oscillation $\omega$ as:

\begin{equation}\label{eq:cond2}
\frac{1}{\abs{\Phi}}\frac{d\abs{\Phi}}{dt}\ll\omega.
\end{equation}

Conditions (\ref{eq:cond1})-(\ref{eq:cond2}) set the so-called fast oscillation regime of the Klein-Gordon equation (\ref{eq:KG}). Following this prescription, (\ref{eq:real}) can be reduce to

\begin{equation}\label{eq:oscrap}
\omega^2=2c^2\frac{dV}{d\abs{\Phi}^2}.
\end{equation}
This allows us to write the fast oscillation condition in terms of the charge $Q$ defined in (\ref{def:Q}), which becomes

\begin{equation}\label{eq:conservacion}
\frac{Q^2\hbar^2c^4}{a^6\abs{\Phi}^4}=2c^2\frac{dV}{d\abs{\Phi}^2}.
\end{equation}

Using the expression for the scalar field potential (\ref{eq:V}), we can approximate (\ref{eq:energia}) using the condition (\ref{eq:oscrap}) as 

\begin{eqnarray}\label{eq:energiared}
\epsilon&=&\frac{1}{2c^2}\left[\left(\frac{d\abs{\Phi}}{dt}\right)^2+\omega^2\abs{\Phi}^2\right]+\frac{m^2c^2}{2\hbar^2}\abs{\Phi}^2-\frac{2\pi A_s m}{\hbar^2}\abs{\Phi}^4 \nonumber \\
&\approx&\frac{m^2c^2}{\hbar^2}\abs{\Phi}^2-\frac{6\pi A_s m}{\hbar^2}\abs{\Phi}^4,
\end{eqnarray}
By a similar approach, the scalar pressure from (\ref{eq:paraintegrar}) can take the approximate form

\begin{equation}\label{eq:Pmaschica}
P\approx-\frac{2\pi A_s m}{\hbar^2}\abs{\Phi}^4.
\end{equation}

Solving  (\ref{eq:energiared}) for $\abs{\Phi}^2$, we obtain two possible branches that correspond to solutions of the Einstein-Klein-Gordon system in the fast oscillation approximation

\begin{equation}\label{eq:phi2}
\abs{\Phi}^2=\frac{c^2m}{12\pi A_s}\left(1\pm\sqrt{1-\frac{24 \pi A_s\hbar^2}{m^3c^4}\epsilon}\right).
\end{equation}
Notice that this is a different result in comparison to the repulsive self-interaction case \citep{Li:2013nal},
where there is an unique branch in the solution since only the $(+)$ sign of the square root is possible. Furthermore, in  \citet{Suarez:2016eez} was shown that for the attractive self-interaction case (\ref{eq:phi2}) and when we take
 the negative sign, the scalar field undergoes a matter-like phase (and even an inflation epoch). While for the positive branch, the solution behaves as dark energy. From this point forward we will take the positive sign, to focus on that particular branch.

Therefore, by using (\ref{eq:phi2}) in (\ref{eq:Pmaschica}) we obtain

\begin{equation}\label{eq:Pchida}
P(\epsilon)=-\frac{m^3c^4}{72\pi A_s\hbar^2}\left(1+\sqrt{1-\frac{24\pi A_s\hbar^2}{m^3c^4}\epsilon}\right)^2.
\end{equation}
Physical solutions of this latter equation
correspond to those values of $\epsilon$ smaller than a certain $\epsilon_i$:
\begin{equation}
\epsilon_i=\frac{m^3c^4}{24\pi A_s\hbar^2}.
\end{equation}

From the two latter expressions, notice that
$P(\epsilon_i)=-\frac{m^3c^4}{72\pi A_s \hbar^2}$, implies that $w_i=\frac{P(\epsilon_i)}{\epsilon_i}=-1/3$.

The scale factor for which the energy density takes the value $\epsilon_i$ can be calculated by inserting the value of $\abs{\Phi}^2$ evaluated in $\epsilon_i$, and taking the result on the fast oscillation condition (\ref{eq:conservacion}):

\begin{equation}\label{eq:ainit}
a_i=\sqrt[3]{\frac{12\sqrt{3}\pi A_s\hbar^2 |Q|}{m^2c^2}}.
\end{equation}
For convenience, we re-define a dimensionless quantity in terms of  the differential equation for the energy density as

\begin{equation}
\bar{\epsilon}=\frac{\epsilon}{\epsilon_i},
\end{equation}
therefore (\ref{eq:paraintegrar}) can be written as

\begin{equation}\label{eq:barepsilon}
\frac{d\bar{\epsilon}}{da}=-\frac{3}{a}\left[\bar{\epsilon}-\frac{1}{3}\left(1+\sqrt{1-\bar{\epsilon}}\right)^2\right].
\end{equation}

Evaluating in $\epsilon=\epsilon_i$, we can see that $d\bar{\epsilon}/da$, takes a negative value of $-\frac{2}{a_i}$, therefore for $a<a_i$ the solution is not valid. The value $a_i$ indicates the scale factor at the time when the scalar field turns on. Furthermore, at $a\rightarrow\infty$, $\epsilon$ approaches to a constant value. 

Now, taking the fast oscillation equation (\ref{eq:conservacion}) and inserting $\abs{\Phi}^2$ from (\ref{eq:phi2}) we obtain

\begin{equation}\label{eq:eccubica}
\left(\frac{a_i}{a}\right)^6=3\left(1+\sqrt{1-\bar{\epsilon}}\right)^2-2\left(1+\sqrt{1-\bar{\epsilon}}\right)^3.
\end{equation}

In order to find the asymptotic value of $\epsilon$, when $a\rightarrow\infty$, we should consider the fast oscillation equation (\ref{eq:conservacion}), which for potential (\ref{eq:V}) takes the form

\begin{equation}\label{eq:lala}
\frac{Q\hbar c^2}{a^3}=\sqrt{2}c\abs{\Phi}^2\sqrt{\frac{m^2c^2}{2\hbar^2}-\frac{4\pi A_s m}{\hbar^2}\abs{\Phi}^2}.
\end{equation}
Since $\epsilon$ decreases with $a$, then $\abs{\Phi}^2$ increases as $a\rightarrow\infty$ as we can notice from (\ref{eq:phi2}), therefore the term inside the square root in (\ref{eq:lala}) should vanish
as $a\rightarrow\infty$, leading to an asymptotic value of

\begin{equation}\label{eq:SFasymp}
\abs{\Phi_\Lambda}^2=\frac{mc^2}{8\pi A_s}.
\end{equation}
Using (\ref{eq:energiared}) and (\ref{eq:Pchida}) we can obtain

\begin{align}
&\epsilon_\Lambda=\frac{m^3c^4}{32\pi A_s\hbar^2}=\frac{3}{4}\epsilon_i,\\
&P(\epsilon_\Lambda)=-\epsilon_\Lambda.
\end{align}
Notice how in the limit $a\rightarrow\infty$, the scalar field has an EoS with a value $w_\Lambda=-1$. Therefore the EoS interpolates between the values $-1/3$ and $-1$. This is a result of both the rapidly oscillating behavior of the field and the chosen peculiar branch, although not a general property of a homogeneous complex cosmological scalar field nor a direct consequence of having a non-zero conserved quantity $Q$. This result is very different from the one that would have been obtained for the other branch of the solution or even for the real case. In those cases we would not have a scalar field solution with $w<0$ that turns on at a certain scale factor $a_i$ and not before.

\subsection{Exact solution for the dark energy term-like}
\label{subsec:exact}

To obtain an expression for $\epsilon$ in terms of the scale factor, we have to solve the equation \ref{eq:eccubica}. This can be obtained making the change of variable 

\begin{equation}
\zeta=\sqrt{1-\bar{\epsilon}}+\frac{1}{2}.
\end{equation}

The latter leads to an expression in terms of a cubic equation
\begin{equation}
\zeta^3-\frac{3}{4}\zeta+\frac{1}{2}\left(\frac{a_i^6}{a^6}-\frac{1}{2}\right)=0,
\end{equation}
which has three real solutions. However, it must satisfy the conditions $\zeta(a_i)=\frac{1}{2}$ and $\zeta(a\rightarrow\infty)=1$. The only solution that satisfy these conditions is 

\begin{equation}\label{eq:zeta}
\zeta(a)=\cos\left[\frac{1}{3}\arccos\left(1-2\frac{a_i^6}{a^6}\right)\right],
\end{equation}
in
terms of this function $\zeta(a)$, the energy density and the EoS parameter are given by the following expressions

\begin{align}\label{eq:epseps}
\epsilon(a)&=\left[1-\left(\zeta(a)-\frac{1}{2}\right)^2\right]\epsilon_i,
\end{align}
\begin{align}
w(a)&=-\frac{\left(\zeta(a)+\frac{1}{2}\right)^2}{3-3\left(\zeta(a)-\frac{1}{2}\right)^2}.\label{eq:eos1} 
\end{align}
This is the so-called Complex Scalar Field Dark Energy (CSFDE) model.
These solutions should be considered only in certain region $a_i<a<a_e$ of the evolution of the Universe, the upper limit $a_e$ is defined as the scale factor when the fast oscillation regime ceases to be valid, which we will calculate below. This is evident from (\ref{eq:conservacion}), since $\omega$ get suppressed by the term $a^6$, while $\abs{\Phi}$ goes to a constant value. From now on, $ a $ will only be referred to this range. First, we must make sure that the solution at $a_i$ satisfy the fast oscillation approximation described in the latter section. 

Under these ideas, the fast oscillation condition $\omega\gg H$ is given by

\begin{equation}
\frac{Q^2\hbar^2c^4}{a^6\abs{\Phi}^4}\gg\frac{8\pi G}{3c^2}(\rho_m+\epsilon),
\end{equation}
where $\rho_m=\Omega_m/\rho_{\text{crit}}$, see (\ref{eq:matter_density}).
By performing the substitution of $\abs{\Phi}^2$ using (\ref{eq:phi2}), re-writing it in terms of $\bar{\epsilon}$ and, finally, taking $\epsilon\gg\rho_m$, we can obtain 

\begin{equation}\label{eq:realcondition}
\left(\frac{a_i}{a}\right)^2\gg\frac{mG}{3c^2A_s}\bar{\epsilon}(1+\sqrt{1-\bar{\epsilon}})^2.
\end{equation}
This condition will be satisfied initially if

\begin{equation}
\frac{3c^2 A_s}{mG}\gg 1.
\end{equation}

In order to compute the value $a_e>a_i$, where the solution is no longer valid, we will consider the end of the fast oscillation regime when $\omega=NH$ (with $N=200$ analogous to \cite{Li:2013nal}). If $a_e\gg 1$ and also $a_e\gg a_i$, in order to be able to make the approximations $\epsilon\gg\rho_m$ and $a_i/a_e\ll1$ in  (\ref{eq:realcondition}), then the end value of the scale factor will be 

\begin{equation}
a_e\approx\sqrt[6]{\frac{\frac{768}{N^2}\pi^2 A_s^3\hbar^4 Q^2}{Gm^5 c^2}}.
\end{equation}

\begin{figure}
    \centering
    \includegraphics[width=0.49\textwidth,height=!,angle=0]{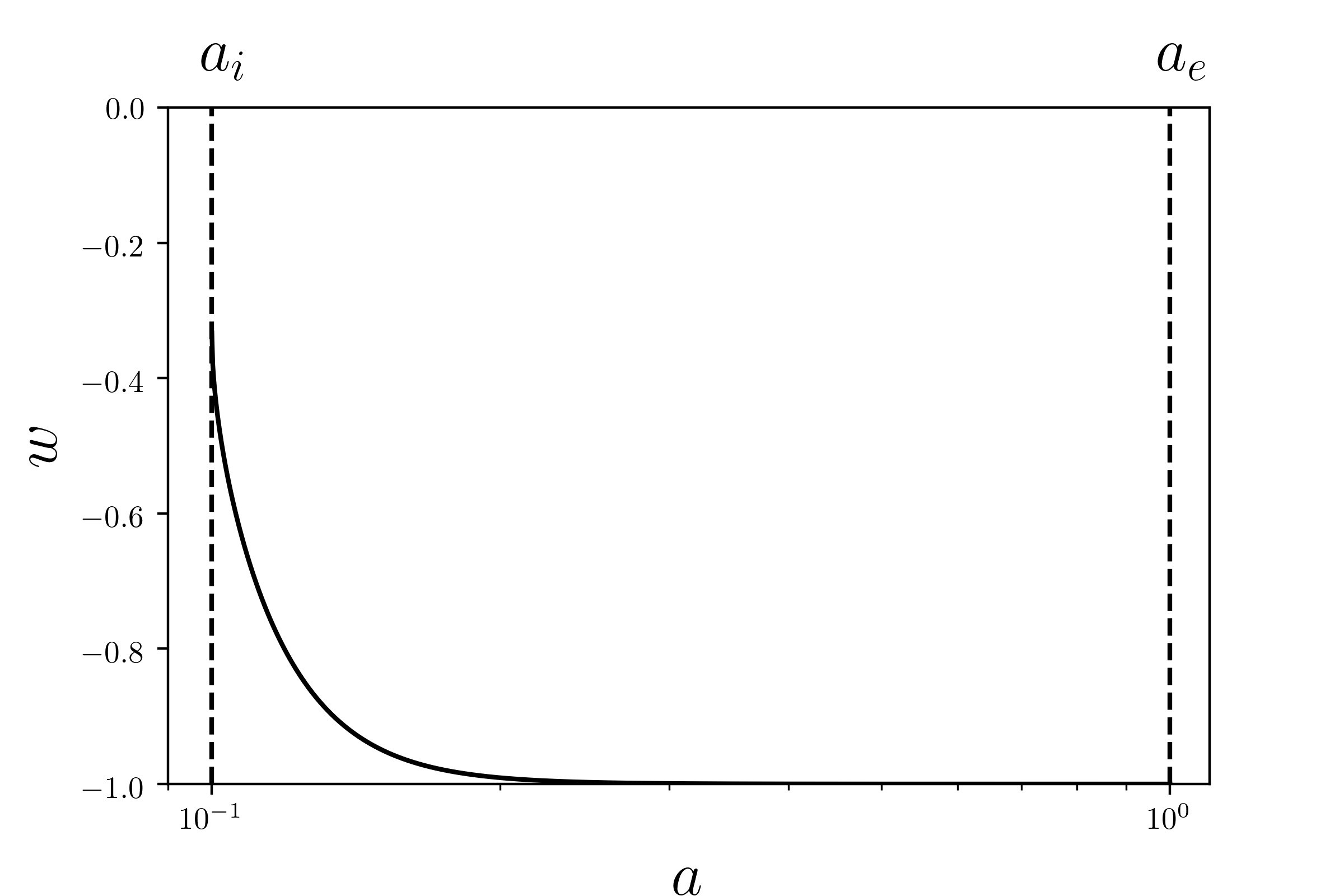}
    \caption{Evolution of the $w$ Eq.~(\ref{eq:eos1}) as a function
    of the scale factor, $a$, in the quintessence model.}
    \label{fig:w}
\end{figure}

In Fig. \ref{fig:w} we show an example for the evolution of the equation of state parameter $w$ between the values for the scale factor $a_i$ and $a_e$, determined by specific values of $m$, $A_s$ and $Q$. In this example we take\footnote{This particular choice of $a_i$ and $a_e$ in our example reduces the dimension of the free parameter space from 3 to 2, thus we can put $Q$ and $A_s$ in terms of $m$: $Q=\frac{4a_{\text{min}}^9mc^4}{27\sqrt{3}\pi NG\hbar^4}$,
$A_s=\frac{9N^2G\hbar^2 m}{16a_{\text{min}}^6c^2}$.} $a_i$ to be the value $a_{\text{min}}=0.1<1/(1+z_{\text{max}})$ where $z_{\text{max}}=2.26$ corresponds to the maximum redshift used in the multiple data sets within the analysis described in the next section. In this way, we ensure that the scalar field is present throughout the $a$ range of the analysis. We have restricted this example to the case where $a_e=1$, thus ensuring that the limit of rapid oscillations and therefore the cosmological constant type behaviour continues to be valid today.

\begin{figure}
    \centering
    \includegraphics[width=0.49\textwidth,height=!,angle=0]{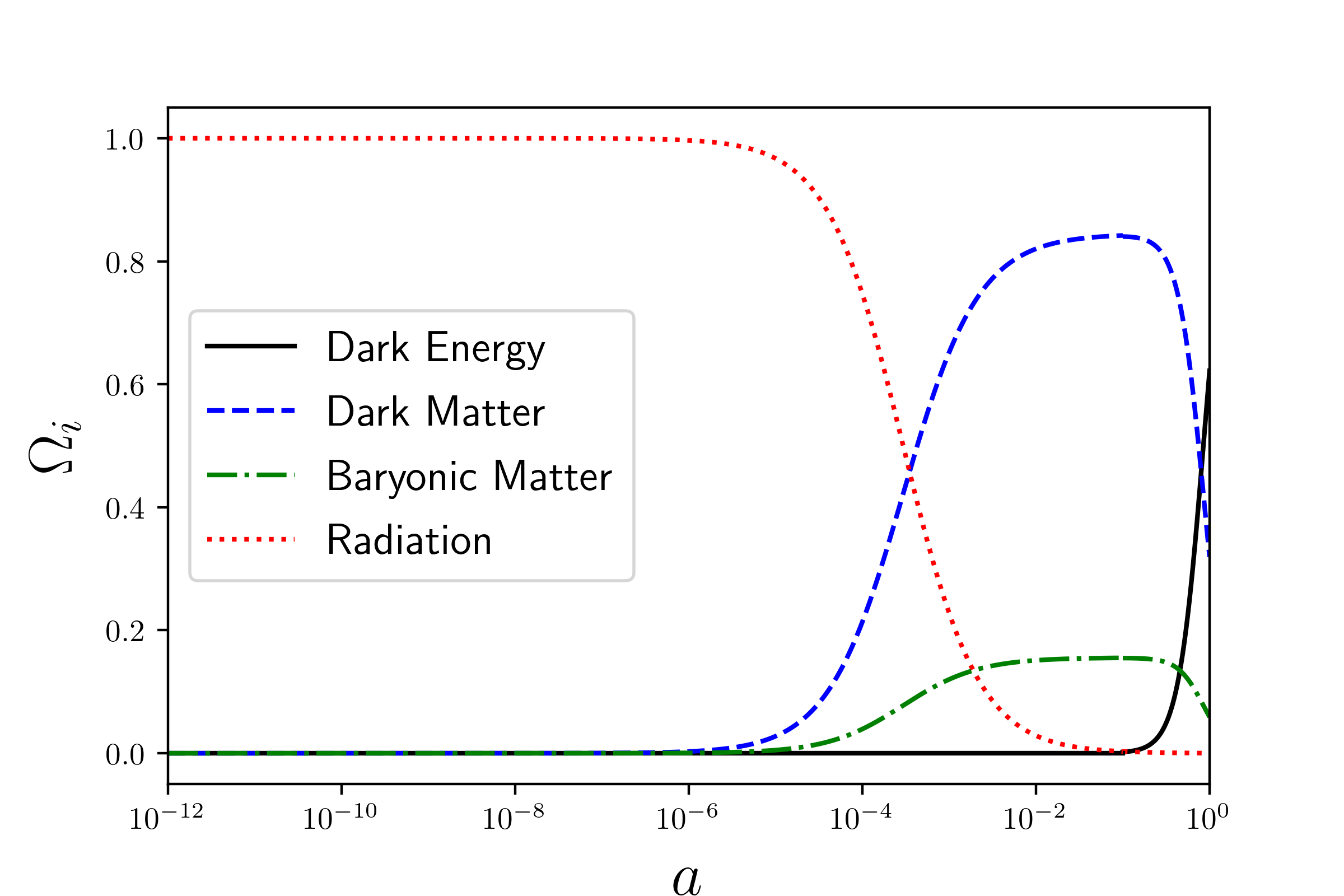}
    \caption{Evolution of the $\Omega_i$ as a function
    of the scale factor, $a$, in the quintessence model.}
    \label{fig:Omegas}
\end{figure}

To give intuition about what is happening in the complete cosmological model where dark energy is described by the scalar field solution described at Sec.~\ref{sec:quintessence}, we present in Fig. \ref{fig:Omegas} the energy density fractions of the quintessence model with the same values $a_i$ and $a_e$ as in Fig. \ref{fig:w}, additionally, for our example we have chosen the initial scalar field energy density to be $\epsilon_i\equiv\frac{4}{3}\epsilon_\Lambda=\frac{3}{4}\rho_{\text{crit}}\Omega_\Lambda$. In other words, we have chosen that the asymptotic value for the energy density of the scalar field coincides with the current energy density for $\Lambda$ in the pivot model. Interestingly, it turns out that this condition on the example fixes the three free parameters of our model, leading to a mass
$m\sim 10^{-22}\text{eV}/c^2$, frequently used in the ultralight models of dark matter \citep{Magana:2012ph,Schive:2014dra,Hui:2016ltb}. The Fig.~\ref{fig:Omegas} is almost indistinguishable from the corresponding figure for the 
$\Lambda$CDM model, this is because the discontinuity for $\epsilon$ appears in an epoch where the contribution to the total energy density of the scalar field is relatively small and also because $\epsilon$ quickly tends to the $\epsilon_\Lambda$ value, as can be inferred from Fig. \ref{fig:w} and equation \ref{eq:epseps}. 
%


\subsection{Parametric Equation of State in the late cosmic acceleration approximation}
\label{sec:eos_scalar}

Let us write explicitly (\ref{eq:eos1}) as an effective dark energy EoS described by a complex scalar field with a Bose-Einstein condensate-like potential.
By using
the standard definition $a=1/(1+z)$ and expand the function $\zeta$ in (\ref{eq:zeta}) with the assumption $a \gg a_i$ we obtain:
\begin{equation}\label{eq:aprox_eosSF}
    w(z)=w_0+w_a(1+z)^6,
\end{equation}
where $w_0=-1$ and $w_a=\frac{16}{27}{a_i}^6$. Notice that this generic expression for the EoS impose directly on $w_0$ the cosmological constant value. We will refer to this particular parametrization of the scalar field model as \textit{parametric form for the CSFDE model}.

The quantity $a_i$ which completely determines $w_a$, is restricted to have values consistent with a scalar field present at any time in the past. Therefore we should take the range $0<a_i<1$ which, translated to $w_a$, corresponds to the range
\begin{equation}\label{eq:wa}
w_a \in \left(0,\frac{16}{27}\right).
\end{equation}

Actually the validity of the parametric equation of state (\ref{eq:aprox_eosSF}) requires pushing the value $a_i$ further back in time. We could take, for instance $a_i=0.3$, which satisfies the above conditions. In this case, the parametric equation of state (\ref{eq:aprox_eosSF}) has a maximum absolute error with respect to the exact case (\ref{eq:eos1}) of $8\times10^{-2}$.

We should remark that (\ref{eq:aprox_eosSF}) is not obtained as in the traditional derivation of the solution of the conservation equation, where an effective dark energy fluid needs to be consider and certain fixed values of $w$ denote the different matter in the universe. 


\section{Observational constraints}
\label{sec:surveys}

To perform the statistical analyses for the parametric CSFDE (\ref{eq:aprox_eosSF}) and to find current constraints of the model, we are going to consider in this paper late-time data sets as SNeIa (Pantheon), Observational Hubble data (OHD) and Baryon Acoustic Oscillations (BAO). 

\begin{figure*}
\centering
\includegraphics[width=0.62\textwidth,origin=c,angle=0]{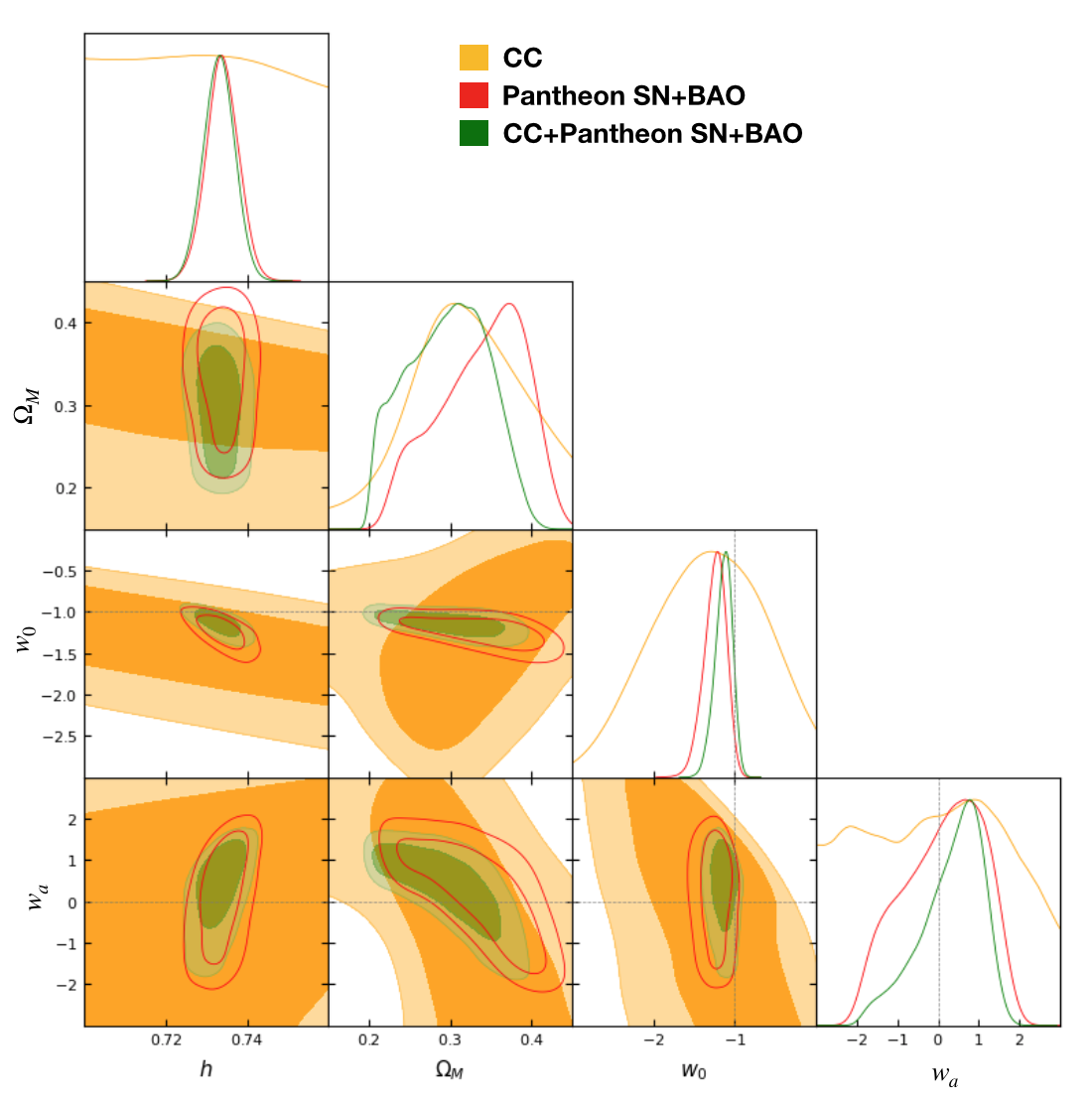}
\caption{The 68\% confidence level (C.L.), 95\% CL. and 99.7\% regions inferring from the parametric CSFDE (\ref{eq:aprox_eosSF}) using CC (yellow C.L), Pantheon supernovae + BAO (red C.L) and the full sample (CC+SN+BAO) (green C.L).}
\label{fig:contour_SF1}
\end{figure*}

Each observational data has the following features:

\begin{itemize}

\item Pantheon SNeIa compilation: This sample is one of the latest Type Ia Supernovae (SN) compilations \citep{Scolnic:2017caz} and  it contains 1048 SNeIa at redshift $0.01<z<2.26$. The constraining power of this kind of supernovae is due that this observation can be used as standarizable candles. This can be implemented through the use of the distance modulus
\begin{eqnarray}
\mathcal{F}(z,\Theta)_\text{theo}=5\log_{10}\left[D_L(z,\Theta)\right]+\mu_0,
\end{eqnarray}
where $D_L$ is the luminosity distance given by
\begin{eqnarray}
D_L(\Theta)=(1+z)\int_0^z{\frac{c\, dz'}{H_0 E(z',\Theta)}},
\end{eqnarray}
and $\Theta =\{w_0,w_a\}$ is the vector with the free cosmological parameters to be fitted. We notice that the factor $c/H_0$ can be absorbed in $\mu_0$. Furthermore, we can write $\Delta\mathcal{F}(\Theta)=\mathcal{F}_{\text{theo}}-\mathcal{F}_{\text{obs}}$, using for this purpose the  distance modulus $\mathcal{F}_{\text{obs}}$ associated with the observed magnitude. At this point it may be thought  that a possible $\chi_{SN}^2$ is given by
\begin{eqnarray}
\chi_{SN}^2(\Theta)&=&\left(\Delta\mathcal{F}(\Theta)\right)^{T}\cdotp C_{SN}^{-1}\cdotp \Delta\mathcal{F}(\Theta),
\end{eqnarray}
where $C_{SN}$ is the total covariance matrix.
This equation can be used to contain the nuisance parameter $\mu_0$, which in turn is a function of the Hubble constant, the speed of light $c$ and the SNeIa absolute magnitude. To circumvent this issue, $\chi_{SN}^2$ is marginalized analytically with respect to $\mu_0$ and we can obtain a new $\chi_{SN}$ estimator 
\begin{eqnarray}
\chi_{SN}^2(\Theta)&=&\left(\Delta\mathcal{F}(\Theta)\right)^{T}\cdotp C_{SN}^{-1}\cdotp \Delta\mathcal{F}(\Theta)+\ln{\frac{S}{2\pi}}-\frac{k^2(\Theta)}{S},\nonumber\\
\end{eqnarray}
where $S$ is the sum of all entries of $C_{SN}^{-1}$. This equation gives an estimation of the precision of these data independently of $\Theta$, and $k$ is $\Delta\mathcal{F}(\Omega_m,\Omega_r,\Omega_\Lambda)$ but weighed by a covariance matrix as follows:
\begin{equation}
k(\Theta)={\left(\Delta\mathcal{F}(\Theta)\right)^{T}\cdotp C_{SN}^{-1}}.
\end{equation}

Also, for this sampler we 
are taking the nuisance parameter $M$ inside the sample, for this we choose the respective values of $M$ from a statistical analysis of the $\Lambda$CDM model with a fixing $H_0$ from the Late Universe measurements (SH0ES + H0LiCOW) as $H_0 =73.8 \pm 1.1 \text{km/s/Mpc}$ with $M=-32.79$.

\item BAO measurements: we consider the sampler of 15 transversal measurements obtained in a quasi model-independent approach. 
This
can be done by computing the 2-point angular correlation function tracers via $D_A (z; r_{\text{drag}})$ \citep{Nunes:2020hzy}. 
The sampler is given in a redshift range $[0.11,2.225]$. 
These kind of observations contribute with important features by comparing the data of the
sound horizon today to the sound horizon at the time of recombination (extracted from the CMB anisotropy~data).
The BAO distances are given by
$d_{z} \equiv \frac{r_{s}(z_{d})}{D_{V}(z)}$, with  $r_{s}(z_{d}) = \frac{c}{H_{0}} \int_{z_{d}}^{\infty}
\frac{c_{s}(z)}{E(z)} \mathrm{d}z$ being the comoving sound horizon at the baryon dragging epoch,
$c$ the light velocity, $z_{d}$ is the
drag epoch redshift and $c^{2}_{s}= c^2/3[1+(3\Omega_{b0}/4\Omega_{\gamma 0})(1+z)^{-1}]$ the sound speed
with $\Omega_{b0}$ and $\Omega_{\gamma 0}$ the present values of baryon and photon density parameters, respectively. The dilation scale is given by
\begin{equation}
D_{V}(z,\Omega_m; \Theta) = \left[   \frac{c
\, z (1+z)^2 D_{A}^2}{H(z, \Omega_m; \Theta)} \right]^{1/3}\,,
\end{equation}
where $D_{A}$ is the angular diameter distance
\begin{equation}
D_{A}(z,\Omega_m; \Theta) = \frac{1}{1+z} \int_{0}^{z}
\frac{c \, \mathrm{d}\tilde{z}}{H(\tilde{z}, \Omega_m;\Theta)} \,,
\end{equation}
where $\Theta=\{w_0,w_a\}$.
Through the comoving sound horizon, the distance ratio $d_{z}$ is
related to the expansion parameter $h$ (defined such that {$H \doteq 100 h$)} and the physical densities $\Omega_{m}$ and $\Omega_{b}$. To connect the BAO data with SNeIa (Pantheon) to CMB data (PL18), we consider the Alcock-Paczynski distortion parameter: 
    \begin{eqnarray}
    F(z,\Theta)=(1+z)\frac{D_{A}(z,\Theta)H(z,\Theta)}{c}.
    \end{eqnarray}
Notice that this possible by calibrating the $D_A$ from BAO with the $d_{L}$from supernovae in a cosmology-independent way and 
we define:
     \begin{eqnarray}
    \chi^2_{BAO}=(\Delta\mathcal{F}_{\textit{BAO}})^{T}\cdotp C_{\textit{BAO}}^{-1}\cdotp\Delta\mathcal{F}_{\textit{BAO}},
    \end{eqnarray}
    where $\Delta\mathcal{F}_{\textit{BAO}}$ is the difference between the observational data and the resulting value for $\Theta$, and $C_{\textit{BAO}}^{-1}$ is the inverse of the covariance matrix reported in the reference mentioned above.

\item  Observational Hubble data (CC): we consider a sample of 51 measurements in the redshift range $0.07< z < 2.0$ \citep{Moresco:2016mzx}.  A calibration of this sample was presented in \citet{Magana:2017nfs}. Moreover, we should be careful since this sample contains data from BAO that can overlapping the sampler. This sample gives a measurement of the expansion rate without relying on the nature of the metric between the chronometer and us as observers.  The normalised parameter $h(z)$ can be compute by considering the values of SH0ES and H0LiCOW given above. In this sample are content 31 data points from passive galaxies and 20 data points are estimated from BAO data under a $\Lambda$CDM prior. However, BAO OHD data points can be computed by using the $r_s$ at the drag epoch from PL18.

To perform the fit of the free parameters of our theoretical setting through the construction of a $\chi_H^2$ as 
\begin{eqnarray}
\chi_H^2=\sum_{i=1}^{51}\frac{\left[H\left(z_i,\mathbf{x}\right)-H_{\textit{obs}}(z_i)\right]^2}{\sigma^2_H(z_i)},
\end{eqnarray}
where $H_{\textit{obs}}(z_i)$ is the observed value at $z_i$, $\sigma_H(z_i)$ are the observational errors, and $H\left(z_i,\mathbf{x}\right)$ is the value of a theoretical $H$ for the same $z_i$ with the specific parameter vector $\mathbf{x}$.

\end{itemize}


\begin{table}
{\centering                     
\begin{tabular}{cccc}                 \hline\hline   Parameters & CC  & Pantheon+BAO& CC+Pantheon+BAO\\ \hline
$h$ & $0.714 \pm 0.071$ & $0.734\pm 0.0040$ & $0.733\pm 0.0038$ \\
$w_0$ & $-1.30\pm 0.72$ & $-1.24\pm 0.15$ & $-1.14\pm 0.12$   \\ 
$w_a$ & $-0.8\pm 2.4$ & $0.13 \pm 1.3$ & $0.33 \pm 0.93$ \\ 
$\Omega_{m }$ & $0.325 \pm 0.094$ & $0.337 \pm 0.072$ & $0.296 \pm 0.047$ \\ 
\hline\hline                              
\end{tabular}   
}
\caption{Background best fits values for Eq.~(\ref{eq:aprox_eosSF}). For CC, Pantheon+BAO and CC+Pantheon+BAO.}
\label{tab:results_total1}                           \end{table}     


\section{Methodology}
\label{sec:methodology}

To proceed with the cosmological precision test of the parametric CSFDE model (\ref{eq:aprox_eosSF}), we compute the $\chi^2$-statistic using each of the observational samplers described. 
Then we find the values of the parameters which minimize each of those individual contributions up to 2-$\sigma$. We repeat the procedure using the total sample, i.e., $\chi^2_{\text{Total}}= \chi^2_{\text{SN}} +\chi^2_{\text{BAO}} +\chi^2_{\text{OHD}}$. 
In Table \ref{tab:results_total1} we report the mean and best fits for the cosmological parameters and the model parameters, $w_0$ and $w_a$, for the join samplers CC+BAO+Pantheon SN.

In Fig. \ref{fig:contour_SF1} we provide the confidence regions, which inform us from a Bayesian point of view on the degree of correlations among the cosmological parameters and the statistical tension between the observables. 
For the full data set combination we draw the contours by choosing two shades of a single colour, and we let the dark and light hues represent the $1\sigma$ and $2\sigma$ regions, respectively.

In the analyzes performed, the posterior distribution of the parameter $w_a$ in the EoS remains unconstrained in the range allowed by the model (see equation \ref{eq:wa}). Results in Fig.~\ref{fig:contour_SF1} and Table \ref{tab:results_total1} indicate that the value of $w_a$ spans the entire validity domain in a $1\sigma$ contour. Nevertheless, the more general parametric model (\ref{eq:aprox_eosSF}), can be in fact constrained.
We emphasize that the analysis has been done using the parametric equation \ref{eq:aprox_eosSF} without necessarily being associated with the model for which we have an exact solution. For this parametric model, the result of the statistical analysis indicates that, the parameters $w_0$ and $w_a$ are constrained. However, it is in relation to the proposed theoretical model that the parameters are not constrained. 


\section{Conclusions}
\label{sec:conclusions}

Complex scalar field theory has been used from a Bose-Einstein condensate point of view to describe the cosmic acceleration observed. This makes possible to construct quintessence--complex scalar field scenarios, which can mimic dark energy effects. 
In this particular backstage, we proposed a study of the peculiar branch solution of the Einstein-Klein-Gordon equations in the fast oscillation regime, where the complex scalar field is modelled as an effective dark fluid. As it is standard, from these field equations it is possible to derive an effective equation of state (\ref{eq:aprox_eosSF}), which is a more general model, here called parametric CSFDE. In this panorama, the cosmological parameters related with the model can be constrained using current observational surveys in order to study epochs where the dark energy (at $z=0$) and dark matter ($z\approx 9$) domination occurs.

Using the join samplers as CC+BAO+Pantheon, the parametric CSFDE model (\ref{eq:aprox_eosSF}) was constrained, taking the values, within 1-$\sigma$, of $w_a=0.33\pm0.93$ and $w_0=-1.14\pm0.12$. Moreover, within the considered $w_a$ range, it is not possible to constrain the model, which  best fit parameters are not consistent with the theoretical scalar field model (equation \ref{eq:wa}). The quantity $w_0$, which is a free parameter in the parametric model, is well constrained in all our analyzes within values consistent with the late cosmic acceleration as well as with the theoretical model, given that it is consistent with the constant value $w_0=-1$ within 2-$\sigma$.

We remark that the CSFDE model has a theoretical restriction for the $w_a$ parameter that is not suitable for a statistical analysis with early-time data, e.g., CMB. However, this limitation could be addressed if, for instance, one considers two scalars fields instead of one. Furthermore, some of the conditions applied, as for example the narrow fast oscillation regime, could be relaxed giving enough freedom, so that both using or not using the CMB data, the best fit parameters could be determined.

Finally, we can see that the CSFDE model cannot reproduce an oscillating behaviour of the EoS associated with a dynamical dark energy. 

This result points out the necessity of more than one canonical scalar field to reproduce viable cosmological scenarios.
Further investigation could require combinations of scalar fields like quintom scenarios or changes in the kinetic term. Also exact solutions for other scalar potentials in the fast oscillation regime could lead to models favored by Bayesian analyzes. This will be reported elsewhere.

\section*{Acknowledgments}

This work was supported in part by DGAPA and PAPIIT UNAM through grants IA100220 and IN110218. CE-R acknowledges the Royal Astronomical Society as FRAS 10147 and networking support by the COST Action CA18108. BC and VJ acknowledge support from CONACyT.

\section*{Data Availability}

The data underlying this article will be shared on reasonable request to the corresponding author.



\bibliographystyle{mnras}
\bibliography{biblio} 



\bsp	
\label{lastpage}
\end{document}